\documentclass[apj]{emulateapj}

\begin{document}

\title{A Grid of MHD Models for Stellar Mass Loss and Spin-down Rates of Solar Analogs}

\author{O. Cohen\altaffilmark{1}, J.J. Drake\altaffilmark{1}}

\altaffiltext{1}{Harvard-Smithsonian Center for Astrophysics, 60 Garden St. Cambridge, MA 02138, USA}
\begin{abstract}

Stellar winds are believed to be the dominant factor in spin down of stars over time. However, stellar winds of solar analogs are poorly constrained due to the challenges in observing them. A great improvement has been made in the last decade in our understanding of the mechanisms responsible for the acceleration of the solar wind and in the development of numerical models for solar and stellar winds. In this paper, we present a grid of Magnetohydrodynamic (MHD) models to study and quantify the values of stellar mass-loss and angular momentum loss rates as a function of the stellar rotation period, magnetic dipole component, and coronal base density. We derive simple scaling laws for the loss rates as a function of these parameters, and constrain the possible mass-loss rate of stars with thermally-driven winds.  Despite the success of our scaling law in matching the results of the model, we find a deviation between the ``solar dipole" case and a real case based on solar observations that overestimates the actual solar mass-loss rate by a factor of 3. This implies that the model for stellar fields might require a further investigation with higher complexity which might include the use of a filling factor for active regions, as well as the distribution of the strength of the small-scale fields.   Mass loss rates in general are largely controlled by the magnetic field strength, with the wind density varying in proportion to the confining magnetic pressure $B^2$.  We also find that the mass-loss rates obtained using our grid models drop much faster with the increase in rotation period than scaling laws derived using observed stellar activity. For main-sequence solar-like stars, our scaling law for angular momentum loss vs.\ poloidal magnetic field strength retrieves  the well-known Skumanich decline of angular velocity with time,  $\Omega_\star\propto t^{-1/2}$, if the large scale poloidal magnetic field scales with rotation rate as $B_p\propto \Omega_\star^2$.
  
\end{abstract}



\section{INTRODUCTION}
\label{sec:Intro}

Current understanding of the winds of low-mass stars with outer convection zones is uncertain due to the fact that these winds cannot be directly observed, except in the case of our own Sun.  Consequently, the mass and angular momentum lost to the wind, which are the main ingredients for understanding the spin-down of stars at from the pre-main sequence onward through main sequence evolution \citep{Parker58,schatzmann62,weberdavis67,mestel68}, remains poorly defined.  Several techniques have been used to indirectly determine stellar mass loss rates of different stellar types. These methods include chemical separation and H$\alpha$ profiles \citep[e.g.,][]{Michaud86,LanzCatala92,Bertin95}, radio observations \citep[e.g.,][]{Abbott80,Cohen82,Hollis85,Lim96,Gaidos00}, observations of X-ray emission due to charge exchange \citep[e.g.,][]{Wargelin01}, gravitational settling of metals from wind accreting white dwarfs in binaries \citep{Debes06}, and Ly$\alpha$ absorption at the edge of the astrosphere \citep[e.g.,][]{Wood02,Wood05,Wood06}. In all methods listed above, several uncertain assumptions have been made on the parameters needed to obtain the mass loss rate. For example, the latter method from which results have been generally adopted in recent years assumes the local pressure, density and velocity parameters of the Interstellar Medium (ISM), which are still quite uncertain \citep{Lallement03,Koutroumpa09}. 

It has been known for a long time that there is a relation between the rotation of the star, its age, and its level of magnetic activity \citep[see, e.g.,][]{Pallavicini81,Pizzolato03,gudel07,Wright11}. Despite this, and neglecting the recent possibility that this relation can be disrupted by the existence of a close-in planet \citep[as supported by both empirical data and modeling work by e.g.,][]{Kashyap08, Pont09, Lanza10, Cohen10b}, there is still no generally accepted relation between these parameters and the stellar mass loss rate. \cite{Cohen11} argued that the level of stellar activity  (i.e., the X-ray and UV luminosity) cannot be used as a proxy for the mass loss rate since it represents the magnetic component of the closed coronal loops.  On the Sun, this component changes by an order of magnitude from low to high solar activity, while the mass loss carried by the solar wind is tied to the magnetic component which is opened up to space by the wind, and it is rather constant through the solar cycle \citep{Cohen11,LeChat12} . 

With the extensive observational uncertainties in determining mass loss rates of low-mass stars, a modeling approach can be taken. Considerable progress has been achieved in recent years in the development of both analytical and Magnetohydrodynamic (MHD) models. In particular, numerical MHD models can capture the interaction between the coronal magnetic field and the coronal and wind plasma. The advantage of such an approach is our ability to control the parameters of the numerical experiment and relate each solution to the particular choice of parameter set. 

\cite{Cohen09} (CO09 hereafter) calculated the effect of latitudinal variation of magnetic spots on the stellar mass-loss rate and spin-down rate using a three-dimensional MHD model. They found that these parameters are highly affected by the distribution of the stellar magnetic field. \cite{Cranmer11} have calculated mass-loss rates of cool stars by including a magnetic filling factor in their model (the percentage of starspots covering the stellar disk). They found a good agreement with observations of large number of systems (under the observational uncertainties mentioned above). Alternatively, in a series of papers,  \cite[][with references therein]{Matt12} performed a series of calculations of stellar mass-loss rates using a two-dimensional, axisymmetric MHD model, which is driven by a spherically symmetric, thermally-driven wind \citep{Parker58}, and using different low-order magnetic topologies. A number of calculations of stellar mass-loss rates of particular systems were recently performed using available Zeeman-Doppler-Imagine (ZDI) maps of the stellar surface magnetic field. The ZDI maps were used to drive the MHD coronal models and to constrain their boundary conditions.  Examples of these calculations are \cite[][AB Doradus]{Cohen10a}, \cite[][V374 Peg]{Vidotto11}, and \cite[][{$\tau$} Boo]{Vidotto12}. 

\cite{Matt12} have recently performed a series of 50 steady-state simulations of stellar coronae in order to estimate angular momentum loss rate as a function of stellar magnetic field strength and rotation rate. Their goal was to derive a semi-analytic formulation for the external torque on the star that includes the magnetic, rotational, and gravitational forces, and their role in controlling the spin-down of stars. 

In this paper, we present a grid of three-dimensional MHD simulations in order to calculate stellar mass-loss rates and angular momentum loss rates as a function of stellar dipole magnetic field strength, stellar rotation period, and coronal base density in a similar manner to \cite{Matt12}. However, as we stress in the next section, our approach is very different, and provides a solution for the stellar corona and stellar wind that is firmly grounded in, and calibrated against, observations of the solar wind and magnetic field topology.

In Section \ref{sec:Model}, we describe the numerical model we use and our approach to calculate the stellar parameters. The results are presented in Section \ref{sec:Results}. In Section~\ref{sec:Discussion}, we discuss the dependence of the stellar loss rates on the different parameters. We also formulate an empirical relation between the stellar loss rates and the parameters we use in the model, and discuss whether a dipole is a good proxy for the stellar magnetic field in the context of stellar mass loss rates.  Finally, we state our conclusions in Section~\ref{sec:Conclusions}.


\section{STELLAR WIND MODELS}
\label{sec:Model}

\subsection{MHD MODEL FOR THE STELLAR WIND}
\label{sec:StellarWind}

In order to produce a grid of solutions for the coronal and wind structure of solar analogs, we follow the approach by CO09 to calculate the mass-loss rate and the angular momentum loss rate. We use the {\it BATS-R-US} MHD code \citep{Powell99} in its version developed for the solar corona \citep{Cohen07}, which solves the set of MHD equations for the conservation of mass, momentum, magnetic induction, and energy. The model is driven by boundary conditions for the radial surface magnetic field, where the {\em initial} condition for the three-dimensional field in the whole domain is potential \citep{altschulernewkirk69}. The acceleration of the stellar wind is done via a source term in the energy equation that is scaled semi-empirically with the topology of the initial potential field and the magnetic flux-tube expansion factor according to the relations found for the solar wind by \citet{wangsheeley90} and \citet{argepizzo00} as implemented by \citet{Cohen07}. The code then solves the MHD equations with this source term until the solution is relaxed to a steady-state, which resembles the observed solar wind structure (when using solar observations to drive the model). A detailed description of the model and the equations it solves can be found in \cite{Cohen09} and \cite{Cohen10a}.

The unique feature that makes this model more "realistic" is the semi-empirical relation between the magnetic topology and the wind speed, as well as the base density. The latter is imposed to resemble the observed density difference between the slow and fast wind, and turns out to be crucial in determining the mass loss rate of each solution. In Section~\ref{sec:Discussion}, we discuss how this density difference in a bi-modal stellar wind is the reason for the discrepancy between a solution for a ``dipole solar case" and a solution based on actual solar minimum (nearly a dipole) data. In other models with an assumed spherical wind (and density) distribution imposed on a certain magnetic field topology, the MHD solution relaxes to a steady-state only due to pressure balance between the wind pressure and the magnetic pressure. However, the stellar wind is assumed to simply ``be there", and no mechanism to drive the wind appears in the equations themselves. In our model, a steady-state is obtained due to the same pressure balance, but also due to a volumetric energy source which is dictated by the magnetic field topology. This way, we obtain a realistic, bi-modal steady-state solution for the wind, with faster wind originating from coronal holes and slower wind coming from the boundaries of the helmet streamers as observed in the solar wind \citep[see, e.g.,][]{McComas07}.

\subsection{A GRID OF MODELS FOR SOLAR ANALOGS}
\label{sec:StellarLossRates}

As noted, the advantage of our model is that it successfully reproduces the solar wind in a realistic manner. By assuming that stellar winds of solar analogs are driven by {\it the same process as the solar wind}, we use the model to calculate different steady-state solutions for a grid of models, where we employ three free parameters: 1) the stellar dipole field strength; 2) the stellar rotation period; and 3) the value of the coronal base density (i.e., the inner boundary condition for the density in the simulation domain). For the stellar magnetic field strength, $B_\star$, we use dipole {\it polar} values of $5,10,50,100,$ and $500$ Gauss; for the stellar rotation period, $\Omega_\star$, we use values of $0.5,2,5,10$ and $25$ days; and for the coronal base density, $n_\star$, we use values of $2e8, 5e8, 1e9,5e9,1e10,$ and $5e10\;cm^{-3}$. 

We point out that the values we use for the base density are significantly lower than the value used in \cite{Matt08}, that formed the basis of the calculations in \cite{Matt12}, and in \cite{Vidotto11}, which were in the range $1e12-1e13\;cm^{-3}$. While these high values were obtained from the literature and very active stars do have traces of plasma at such high densities \citep[e.g.,][]{Ness04,Testa04}
these measurements are for plasma in closed coronal loops, where the density is higher. \citet{Testa04} also notes that the filling factor --- the surface area covered by such high density plasma --- is only of the order of a percent or so.  The ambient coronal density, which is the actual source for the wind that leaves the star along {\it open} field lines, is much lower, unless the lower boundary of the numerical model is set at the chromeshere, which is not the case in any of the models discussed here. To support this argument, we show that we obtain the solar mass-loss rate, $\dot{M}_\odot=2-4\cdot10^{-14}\;M_\odot\;yr^{-1}$, for typical solar parameters and a base density value at the lower end of our density range, whereas solar active regions that dominate the X-ray emission tend to have plasma densities in the range $1e9$--$1e10\;cm^{-3}$ \citep[e.g.][]{Doschek97,DelZanna03,Young09}.

For each {\it BATS-R-US} model solution, we calculate the {\it non-spherical and arbitrary} Alfv\'en surface at which the flow speed, $u$, equals the Alfv\'en speed, 
\begin{equation}
u_A=B/\sqrt{4\pi \rho}, 
\label{Alfvenspeed}
\end{equation}
where $B$ is the magnetic field strength and $\rho=nm_p$ is the mass density ($n$ is the number density and $m_p$ is the proton mass).  Beyond this surface, the wind velocity exceeds the Alfv\'en speed, and the wind is no longer in contact with the star via magnetic fields.  The Alfv\'en surface then represents the surface at which the wind effectively escapes and exerts no more torque on the star.  We then integrate the mass flux, $\rho\mathbf{u}$, over all the Alfv\'en surface elements, $\mathbf{da}_i$. 
This procedure allows us to obtain the total mass-loss rate, $\dot{M}$:
\begin{equation}
\dot{M}=\frac{\partial M}{\partial t}=\int \nabla\cdot (\rho \mathbf{u})dV=\int \rho \mathbf{u}\cdot da=\sum\limits_{i}\rho_i \mathbf{u}_i\cdot\mathbf{da}_i.
\label{Mdot}
\end{equation}
In a similar manner, we integrate the differential version of the formula for the angular momentum loss-rate (Eq. 35 in Weber \& Davis 1967) to obtain the total angular momentum loss rate, $\dot{J}$:
\begin{equation}
\dot{J}=\frac{2}{3}\Omega_\star \dot{M}r_A^2=\frac{2}{3}\sum\limits_{i} \Omega_\star \sin^2{\theta}_i\;r^2_{i} \;\rho_i \mathbf{u}_i\cdot\mathbf{da}_i.
\label{Jdot}
\end{equation}
Here, $r_i$ is the Alfv\'en radius for a particular magnetic field line, or the distance as measured from the center of the star to a given point on the Alfv\'en surface. The $\sin{\theta}_i$ term is due to the fact that the torque applied on the star depends on the angle between the lever arm and the rotation axis. 

Note that in Eq. 3,4, $r_i$ is not constant but is the differential distance at each point of integration over the Alfv\'en surface. The fact that we integrate the differential $r_i$'s allows us to remove any assumption of geometrical symmetries and to perform calculations for any complex topology.


\section{RESULTS}
\label{sec:Results}

\subsection{THE SHAPE AND SIZE OF THE ALFV\'EN SURFACE}
\label{sec:AS}

Eq.~\ref{Mdot} and \ref{Jdot} tell us that the mass and angular momentum loss rates depend on the interplay between the actual mass flux through the Alfv\'en surface (dictated by the density, CO09), and the size of the surface. The size of the surface determines both the total area of integration, as well as the size of the lever arm that applies a torque on a star to spin it down. This is why determining the stellar angular momentum loss rate for a complex field topology is not trivial (CO09). 

\begin{figure*}[h!]
\centering
\includegraphics[width=6.5in]{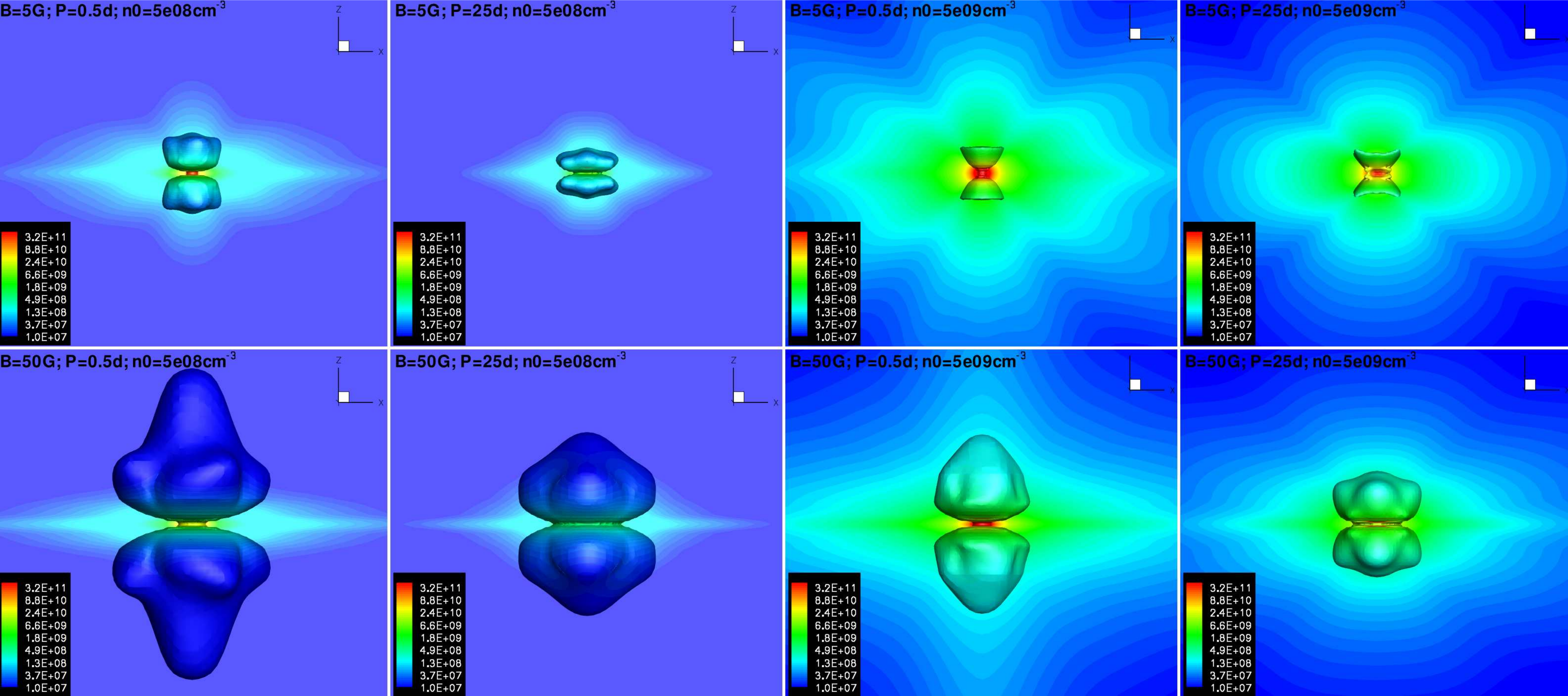}
\caption{Meridional cuts and the Alfv\'en surface in the simulation domains for selected cases (the particular parameters are shown at the top of each panel). Color contours are of the local value of the mass-loss rate.}
\label{fig:f1}
\end{figure*}

Figure~\ref{fig:f1} shows the Alfv\'en surface and a meridional slice of the simulation domain for selected test cases. The color contours represent the local distribution of the mass-loss rate. It can be seen that the size of the Alfv\'en surface increases with an increase of the magnetic field and decreases with the increase of the base density---this is a direct consequence of the dependence of the Alfv\'en speed on these quantities (Eq.~\ref{Alfvenspeed}) and the distance from the star at which the Alfv\'en speed is exceeded by the wind. It can also be seen that the overall mass-loss rate increases with increasing base density. This is partly a consequence of mass continuity in that the base density essentially defines the lower wind boundary condition.  The effect of the rotation period on the Alfv\'en surface is mostly geometrical. When moving from slow to fast rotation, the Alfv\'en surface becomes stretched towards the poles, so that the total surface area increases. 

We note that some of the Alfv\'en surfaces in Figure~\ref{fig:f1} are not fully rotationally symmetric. This is due to artifacts originating from the use of a Cartesian grid of finite resolution for models with rotational symmetry.  We have performed a number of test runs for representative cases with much higher resolution that resulted in rotationally symmetric shapes of the Alfv\'en surfaces. However, comparisons of  mass and angular momentum loss rates from these high-resolution solutions with those of the lower resolution calculations have shown differences of not more than 5 percent. Due to the large number of simulations required for the study and the currently very high computational cost of higher resolution calculations, we have retained the lower resolution approach and consequently some of the results presented here may display residual asymmetric artifacts.

\begin{figure*}[h!]
\centering
\includegraphics[width=6.in]{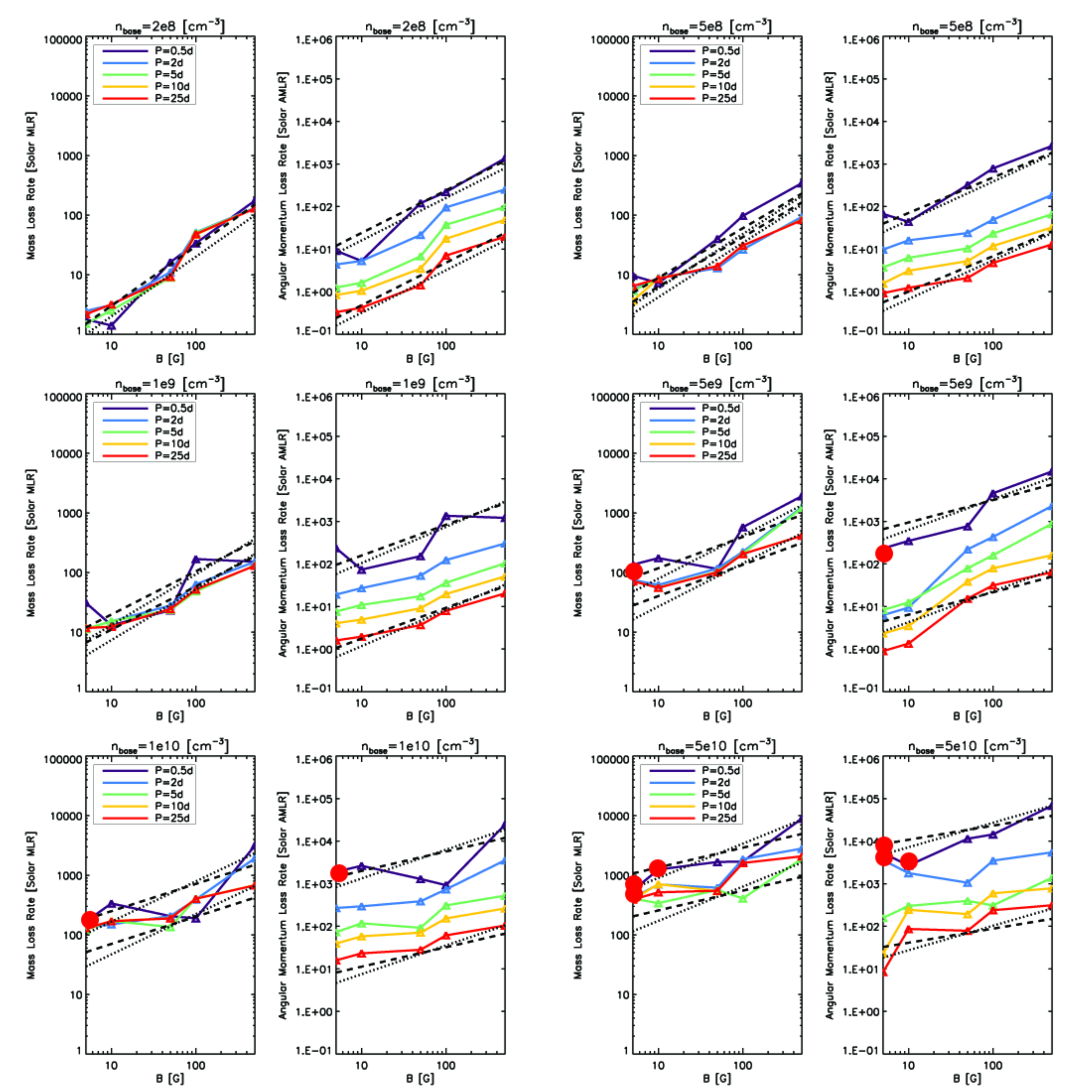}
\caption{The mass-loss rate and angular momentum loss rate as a function of the stellar dipole field strength. Each pair of plots from top to bottom are for a particular value of base density. Each curve in the plots is for a particular rotation period. The red dots mark unphysical cases with the Alfv\'en surface located inside the star. Dashed lines show the scaling laws from Eq.~\ref{MdotScaling} and \ref{JdotScaling} for rotation periods of 25 days (bottom) and 0.5 day (top) with $K=3$, $\alpha=0.8$, $\beta=0.2$, and $\gamma=0.1$. Similarly, the dotted lines show the same for $K=2$, $\alpha=0.8$, $\beta=0.1$, and $\gamma=0.1$.}
\label{fig:f2}
\end{figure*}

Figure~\ref{fig:f2} summarizes the results of our grid of models. Each pair of plots shows the mass-loss rate and the angular momentum loss rate as a function of magnetic field strength for a given base density value. Each curve on the plots represents a particular stellar rotation period. A small number of test cases for very high density and low magnetic field values are considered unphysical because the equatorial rotation velocity on the stellar surface is already greater than the Alfv\'en speed (parts of the Alfv\'en surface are inside the star). These cases are marked with a red spot in the plots of Figure~\ref{fig:f2}. 

\subsection{THE EFFECT OF EACH PARAMETER ON THE STELLAR MASS-LOSS RATE}
\label{sec:parametersImpact}

The most notable trend is the rather linear (in log space, i.e., power law) dependence of the mass and angular momentum loss rates on the dipole field strength. The slope of this trend decreases as we increase the base density, and the difference between the mass loss rate for the weakest and strongest field strength changes from about two orders of magnitude for a base density of $2e8\;cm^{-3}$ to about one order of magnitude for base density of $5e10\;cm^{-3}$. A linear dependence of the angular momentum and mass loss rates on the base density explicitly appears in Eqs.~\ref{Mdot} and \ref{Jdot}. 

\begin{figure*}[h!]
\centering
\includegraphics[width=4.5in]{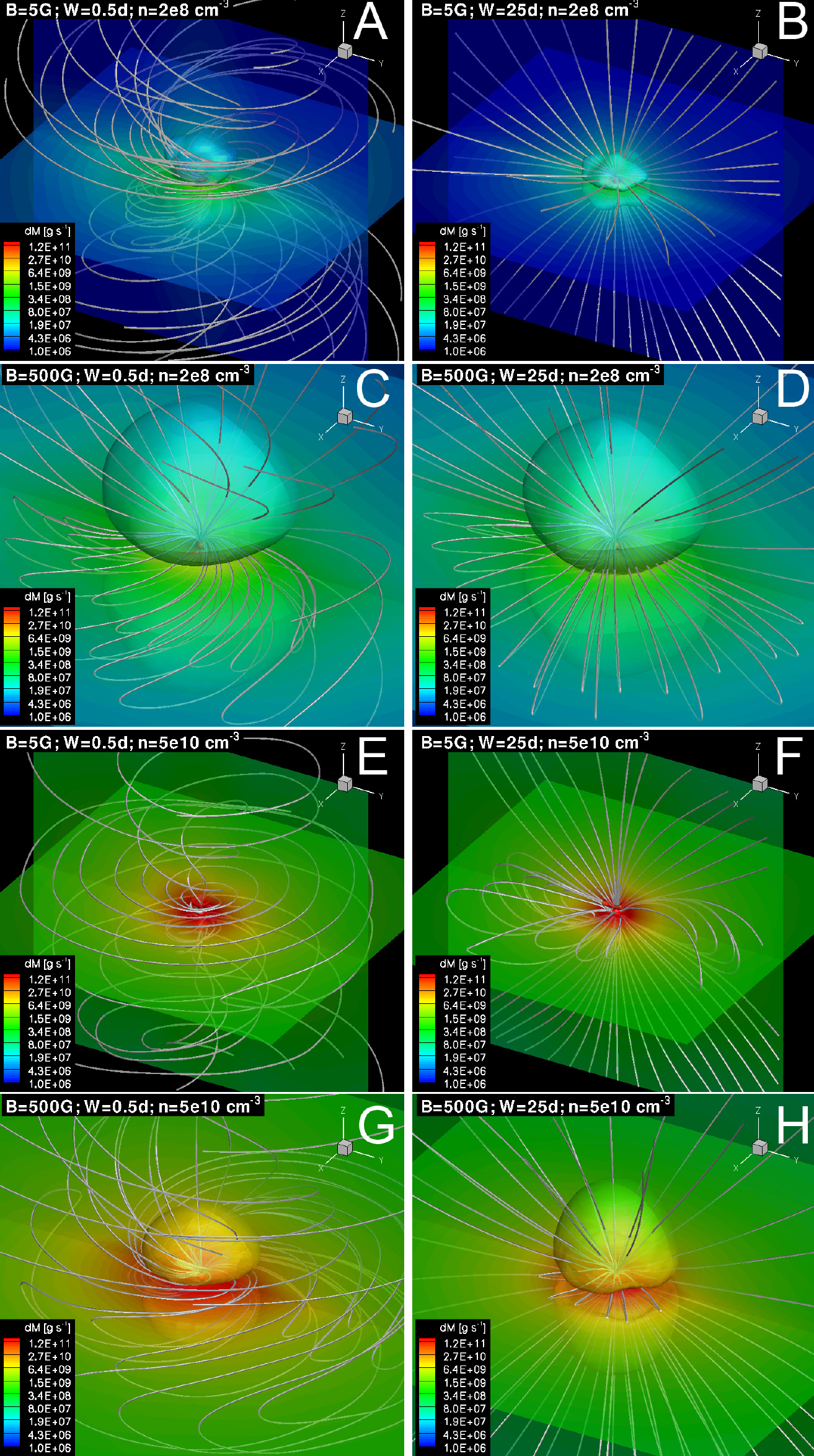}
\caption{Meridional and equatorial cuts, as well as the Alfv\'en surface colored with the local mass-loss rate for cases A-H from Table~\ref{table:t1}. Selected magnetic field lines are also shown as white lines.}
\label{fig:f3}
\end{figure*}

To illustrate the effect of each parameter, Figure~\ref{fig:f3} shows the Alfv\'en surface, the local mass-loss rate distribution, and selected magnetic field lines for the most extreme cases (lowest and highest values for each of the parameters). The parameters of these cases are summarized in Table~\ref{table:t1}. While we consider case E unphysical, we still find the display of the solution valuable for understanding the general trends of the solutions.

\begin{figure*}[h!]
\centering
\includegraphics[width=6.5in]{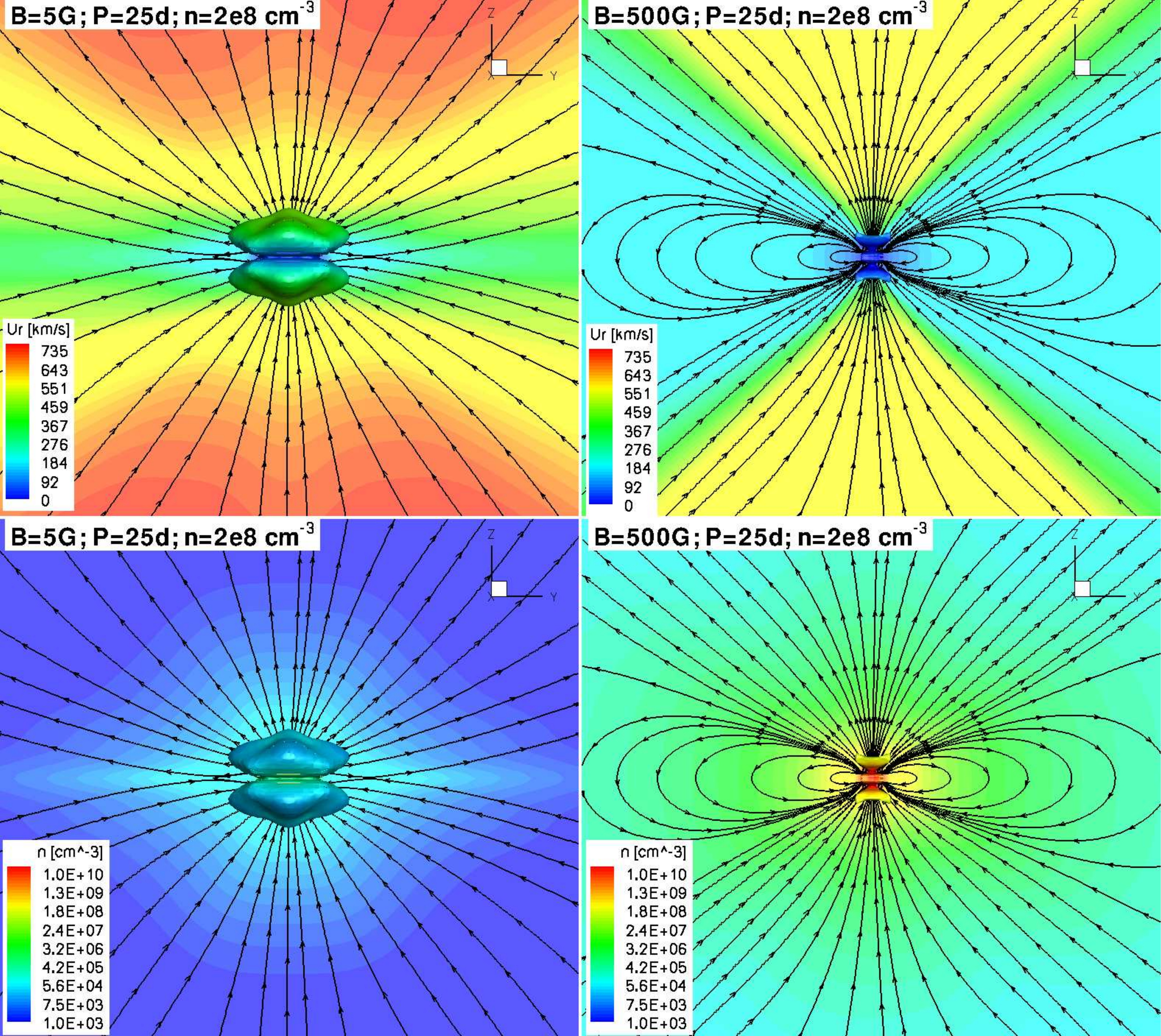}
\caption{Meridional cuts and the Alfv\'en surface for cases B (left) and D (right) colored with radial speed (top) and number density (bottom). Also shown selected magnetic field lines (black)}
\label{fig:f4}
\end{figure*}

First, it can be seen that for slow stellar rotation, the coronal magnetic field lines are stretched and opened up radially by the stellar wind. As we move to very fast rotation, with periods of about 2 days or shorter, the coronal field is wrapped in the azimuthal direction as the azimuthal component of the wind velocity increases with respect to the radial component. For low base density, this tangling does not affect the mass-loss rate much. However, for a strong magnetic field and high base density, the tangling leads to an increase in the coronal density due to the capture of dense plasma in large tangled coronal loops (Case G is about an order of magnitude higher than Case F near the equator). Second, the increase in magnetic field strength modifies the distribution of the stellar wind speed and, as a result, it also modifies the distribution of the coronal density.  The combination of these dictates the mass-loss rate. Figure~\ref{fig:f4} shows meridional slices of cases B and D colored with the contours of the stellar wind radial speed and number density, along with selected magnetic field lines. It can be seen that for a low magnetic field strength of $5\;G$, a fast, less dense stellar wind occupies the majority of the domain except for a narrow region near the equator. The low coronal density is due to the fact that most field lines are open so the stellar wind plasma can escape. In contrast, when the stellar field is strong, a large amount of the magnetic flux remains closed, and only weak and slow wind is developed at lower latitudes. Even at higher latitude, the wind does not exceed a speed of about $500\;km\;s^{-1}$, which is still considered "slow" and dense in terms of the fast solar wind.  For stronger fields, then, the expansion of the magnetic flux tubes that carry the mass flux from the surface into space is greater than for the cases with weaker fields, and the overall wind speed is dominated by the slow, more dense plasma \citep{wangsheeley90,argepizzo00}. As a result, the overall mass-loss rate is greater as B increases.


\section{DISCUSSION}
\label{sec:Discussion}

The trends discussed in Section~\ref{sec:parametersImpact} are directly due to the empirical wind speed--flux tube expansion relation implemented in our model \citep{wangsheeley90,argepizzo00, Cohen07}, which determines the response of the stellar wind {\it acceleration} to the stellar magnetic field. This leads to a non-uniform coronal density based on the stellar wind distribution, and this distribution dictates the mass-loss rate to the wind. As a result, the shape and size of the Alfv\'en surface (which dictate the angular momentum loss rate) are complicated. Even for the simple dipolar case investigated here, the size of the Alfv\'en surface varies strongly with latitude and cannot be used as a single parameter to determine the angular momentum loss rate. 

Our results are consistent with Matt et. al (2012) in that the Alfv\'en surface size increases with increasing magnetic field strength and decreases with increasing base density (see the integrated mass-loss rate in Matt et al. 2012), since by definition, the Alfv\'en surface (or the Alfv\'en speed) depends on the ratio between the magnetic field and the density. In the case of faster rotation, we find that the size of the Alfv\'en surface does not change much in the equatorial regions, but the surface gets stretched outwards and is enlarged in the polar regions. This behavior is also seen in Figures 1 and 2 of Matt \& Pudritz (2008), and it is due to the azimuthal stretching of the field, leading the field strength to drop off with radial distance more slowly than for the case of a purely radial field.  As a result, the field is stronger at a given radial distance and the Alfv\'en radius is pushed outward as noted by Cohen et. al (2010a, 2010b). 

\subsection{SCALING LAWS FOR MASS-LOSS RATE AND ANGULAR MOMENTUM LOSS RATE}
\label{Dipole}

In order to quantify our results, we obtain a scaling law for the stellar mass-loss rate, $\dot{M}_\star$, and the stellar angular momentum loss rate, $\dot{J}_\star$, as a function of the stellar base density, stellar dipole field, and stellar rotation period. These scaling laws are as follow:

\begin{equation}
\frac{\dot{M}_\star}{\dot{M}_\odot}=  K \left( \frac{n_\star}{n_\odot} \right)^\alpha    \left( \frac{B_\star}{B_\odot} \right)^{\left( \frac{n_\odot}{n_\star} \right)^\beta} 
\left( \frac{P_\odot}{P} \right)^{\left(1-\frac{n_\odot}{n}  \right)^\gamma}
\label{MdotScaling}
\end{equation}

\begin{equation}
\frac{\dot{J}_\star}{\dot{J}_\odot} = \left( \frac{P_\odot}{P_\star}\right) \left( \frac{\dot{M}_\star}{\dot{M}_\odot} \right)
\label{JdotScaling}
\end{equation}

We use $\dot{M}_\odot=3\cdot10^{-14}\;M_\odot\;yr^{-1}$ and $\dot{J}_\odot=2\cdot10^{29}\;g\;cm^2\;s^{-2}$, which are the average solar mass-loss rate and angular momentum loss rate between solar minimum and solar maximum periods, $n_\odot=2\cdot 10^8\;cm^{-3}$, $B_\odot=10\;G$, and $P_\odot=27$ days. For lower values of the base density and for the solar case, the above scaling laws are reduced to:

\begin{equation}
\frac{\dot{M}_\star}{\dot{M}_\odot}=  K \left( \frac{B_\star}{B_\odot} \right)
\label{MdotScaling1}
\end{equation}

\begin{equation}
\frac{\dot{J}_\star}{\dot{J}_\odot} = \left( \frac{P_\odot}{P_\star}\right) K \left( \frac{B_\star}{B_\odot} \right)
\label{JdotScaling1}
\end{equation}
As stated in Section~\ref{sec:StellarLossRates}, based on our simulations of the Sun, we believe that the coronal base density should be at the lower end of the density range investigated here.  Unlike the study of \cite{Matt12}, since we have only investigated solar mass stars,  our scaling law does not depend on surface gravity nor the solar mass.

The scaling laws for rotation periods of 0.5 and 25 days are displayed in the plots of Figure~\ref{fig:f2} as dashed and dotted lines, respectively. The set of parameters that provides the best fit for all curves, simultaneously, is $K=3$, $\alpha=0.8$, $\beta=0.2$, and $\gamma=0.1$.  We also show a secondary best fit with $K=2$, $\alpha=0.8$, $\beta=0.1$, and $\gamma=0.1$. In general, these equations describe the dependence of the stellar loss rates on the magnetic field strength, with a slope that is reduced with increasing density, and with the modification caused by rotation. The deviations from these lines are not more than a factor of two for most points, except for the cases with very high (and probably unrealistic) base density, or for the cases of very weak magnetic field, for which the Alfv\'en surface gets very close to the stellar surface. 

\begin{figure*}[h!]
\centering
\includegraphics[width=6in]{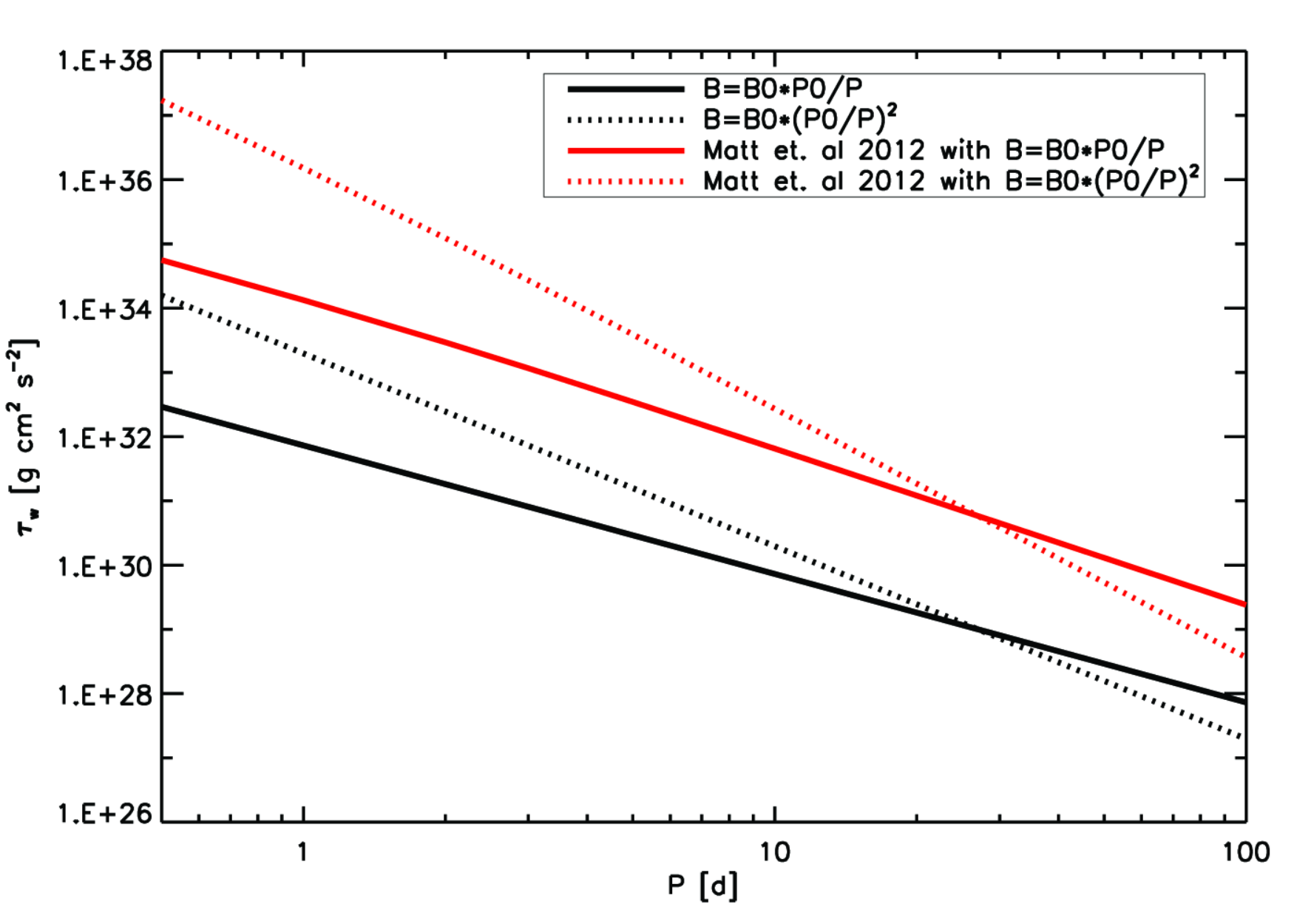}
\caption{The stellar torque as a function of rotation period based on Eq.~\ref{JdotScaling1} (black) for $B\propto \Omega$ (solid) and $B\propto\Omega^2$ (dashed), and based on Eq. 9 in \cite{Matt12} (red) with similar scaling for $B(\Omega)$. The mass loss rate term, $\dot{M}$, in the scaling of \cite{Matt12} is obtained using Eq.~\ref{MdotScaling1} above.}
\label{fig:f5}
\end{figure*}

In Figure~\ref{fig:f5}, we directly compare our scaling law with the scaling law in Eq. 9 from \cite{Matt12}. We also compare the two scaling laws for the magnetic field depending on rotation as $B\propto P^{-1}$ or $B \propto P^{-2}$.  For each rotation period, we obtain the corresponding value of the magnetic field, and then obtain the torque from Eq.~\ref{JdotScaling1} (assuming a base density of $n_\odot$). We also use the corresponding magnetic field to estimate $\dot{M}$ from Eq.~\ref{MdotScaling1}, and use these values of $B(P)$ and $\dot{M(B)}$ to estimate the torque via Eq. 9 from \cite{Matt12}. We use the same values described in \cite{Matt12} for $K1$, $K2$, and $m$, along with the solar rotation, $\Omega_\star=2.7\cdot 10^{-6}$, the solar radius, $R_\star=R_\odot$, and $V_e=600\;km\;s^{-1}$. Overall, the trends of the torques using the two scaling laws are quite similar. However, the overall torque in \cite{Matt12} is about 50 times higher for a field strength of $10G$ and 25 times higher for a field strength of $5G$. Looking at the equation for the torque as a function of the average Alfv\'en radius:
\begin{equation}
\tau_w=\dot{M}\Omega_\star r^2_A,
\label{TorqueScaling}
\end{equation}
and adopting a solar mass loss rate of $2\cdot 10^{-14}\; M_\odot\;yr^{-1}$ ($1.8\cdot10^{12}\;g\;s^{-1}$), and $\Omega_\odot=2.7\times 10^{-6}\;s^{-1}$, our predicted torque is obtained for $r_A=2-10R_\odot$, while the torque predicted by \cite{Matt12} is obtained for $r_A=15R_\odot$, despite of the fact that Table 1 in that paper predicts $r_A$ to be between $8-10R_\odot$ for the solar parameters of $f=0.004$ and $\Upsilon=10^2-10^3$. It is possible that the different in the torque between our scaling and that in \cite{Matt12} is simply based on a different value assumed for the solar case. 

We note that the angular momentum loss rate, $\dot{J_\star}$ is well-represented by a simple scaling of the mass loss rate proportionally to the rotation rate.  This differs from the simple analytical relation 
developed for spherically-symmetric mass-loss that also includes a term involving the Alfv\'en radius, $r_A$, to some power depending on the field geometry \citep[e.g.][]{weberdavis67,Mestel84,Kawaler88,Wood04,Reiners12}.  As we emphasized earlier, for our 3D models $r_A$ is a more complex 3D entity, and more properly the average radius of a local wind density-weighted surface as defined in Eqn.~\ref{Jdot}.
Nevertheless, what we find is that this Alfv\'en ``radius" term is essentially constant in our angular momentum loss results, at least to the precision to which Eqn.~\ref{JdotScaling} represents them.   This might be somewhat surprising given the large range in magnetic field strength investigated, but it has an interesting implication.   The magnetic field at the Alfv\'en radius, $B_{r_A}$ can be expressed in terms of the surface magnetic field, $B_0$, as 
\begin{equation}
B_{r_A}=B_0\left(\frac{R}{r_A}\right)^b,
\label{e:BrA}
\end{equation}
where the index $b$ depends on the magnetic field configuration.  Combining this with Eqn.~\ref{Alfvenspeed}, we can write 
\begin{equation}
u_A=\frac{B_0}{\sqrt{4\pi\rho_A}}\left(\frac{R}{r_A}\right)^b
\end{equation}
where $R$ is the stellar radius and $\rho_A$ is the wind density at the Alfv\'en radius. Since in our models $r_A$ is invariant, we find 
\begin{equation}
\rho_A u_A^2\propto B_0^2,
\end{equation}
that is, the kinetic energy density of the escaping wind responsible for the angular momentum loss is proportional to the magnetic field energy density.  We also find that the wind speed does not vary strongly between models, with a maximum of about 800~km~s$^{-1}$, similar to the solar wind speed and the escape velocity. The bulk of the kinetic energy variations between models are instead due to wind density variations, and so for the {\it equatorial regions that dominate the angular momentum loss}, we can write, very approximately,  $\rho_A \propto B_0^2$.  This simple dependence arises because, in order to escape through the magnetic field, the wind pressure must first overcome the magnetic pressure, and in this way the field pressure near the equator, $B^2/8\pi$, acts as a density regulator.

\subsection{DOES A DIPOLE APPROXIMATION MAKE A GOOD PROXY FOR THE STELLAR MAGNETIC FIELD?}
\label{Dipole}

\begin{figure*}[h!]
\centering
\includegraphics[width=6.5in]{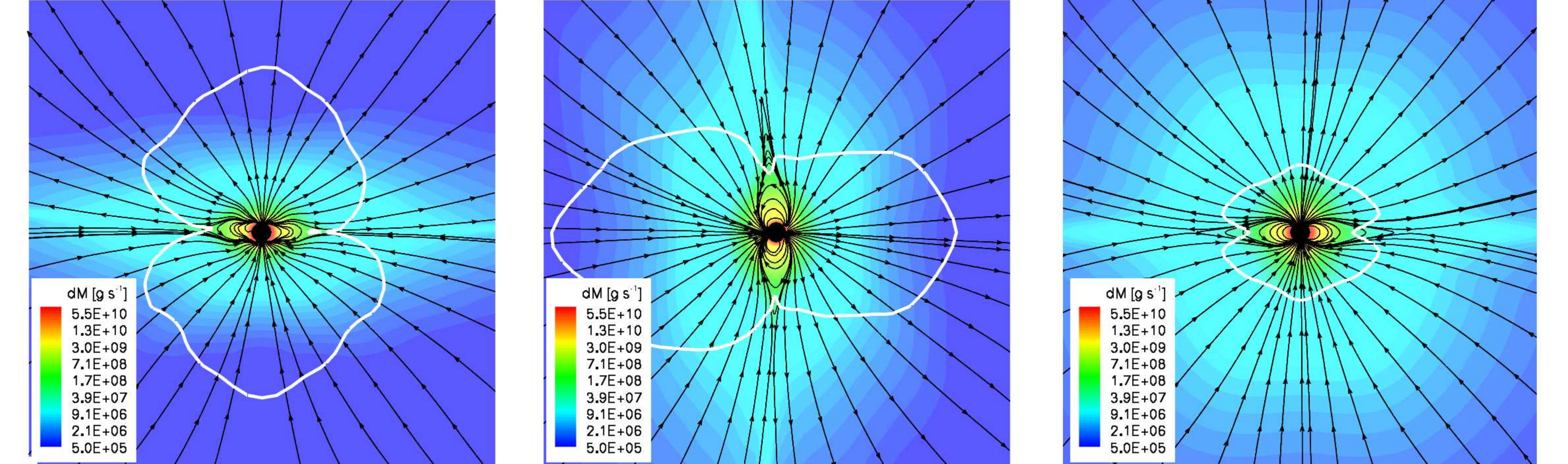}
\caption{Meridional cuts showing the Alfv\'en surface (white solid line), selected magnetic field lines (black lines), and color contours of the local mass-loss rate for solar minimum (left), solar maximum (middle), and a $10\;G$ dipole (right) cases.}
\label{fig:f6}
\end{figure*}

CO09 have shown how active regions on the stellar surface can modify the mass loss and angular momentum loss rates.  We should therefore ask ourselves if a dipole approximation is a good proxy for the stellar field. In order to answer this question, we perform a similar calculation of the mass-loss and angular momentum loss rates but this time, we use actual solar magnetograms to drive the model. These MDI magnetograms\footnote{\tt http://sun.stanford.edu/synop/}  were obtained during a solar minimum period (Carrington Rotation 1922, May 1997) where the solar magnetic field was nearly dipolar, and solar maximum period (Carrington Rotation 1962, April 2000). Figure~\ref{fig:f6} shows a meridional cut colored with contours of local mass-loss rate distribution for the solar minimum and solar maximum cases, as well as a case with a dipole field of $10\;G$. The polar field for the solar minimum case is roughly $8.5\;G$ which is not far from that of the dipole case. The rotation period for all three runs is $P=25$ days and the base density is $n_0=2e8\;cm^{-3}$.

It can be clearly seen that the solar minimum and dipole cases are similar in orientation, but the Alfv\'en surface of the dipole case is much smaller than the one of the real Sun. The reason for this deviation is, again, the dependence of the wind structure on the flux-tube expansion. The expansion in reality in the polar regions seems to be much higher than in the dipole case. As a result, coronal polar regions in the dipole case are occupied by slow, more dense wind than the fast, less dense wind in the real case. It is also possible that for the real case, non-dipole components of the field contribute to the strength of the coronal magnetic field, leading to an increase in the size of the Alfv\'en surface. 

Overall, the dipole case seems to overestimate the solar minimum mass-loss rate, as well as the angular momentum loss rate by a factor of about 3.

\subsection{COMPARISON WITH OTHER STUDIES}

It is well-established, on both theoretical and observational grounds, that the magnetic activity level of a star of a given mass is primarily dependent on the rotation period.  We can derive the expected mass and angular momentum losses as a function of the rotation period if we can assume, or make reasonable guesses for, the values of the other two free parameters in our models, the magnetic field strength and base density. As noted in Sect.~2.2, the relevant magnetic field for driving the wind is the large-scale dipolar field, rather than the smaller scale and much stronger magnetic field associated with active regions and sunspots \citep[see also][]{Cohen11}. From the perspective of understanding stellar winds, this is unfortunate because this dipolar, or poloidal, field is essentially never observed in late-type stars.  Instead, stellar measurements of magnetic fields tend to be dominated by the closed field regions more analogous to solar active regions, where the magnetic fields are up to two orders of magnitude stronger \citep[see also the discussion of][]{Cranmer11}.  While previous studies have utilized these measurements of surface magnetic flux to infer wind properties and relations \citep[e.g.][]{ivanovataam03,Schrijver03b,HolzwarthJardine07,Cranmer11}, our model is somewhat different in that is it explicitly based on the dipolar field component that is expected to accompany, but be distinct from,  the smaller-scale active regions.  We therefore do not follow the same approach.

\begin{figure*}[h!]
\centering
\includegraphics[width=6in]{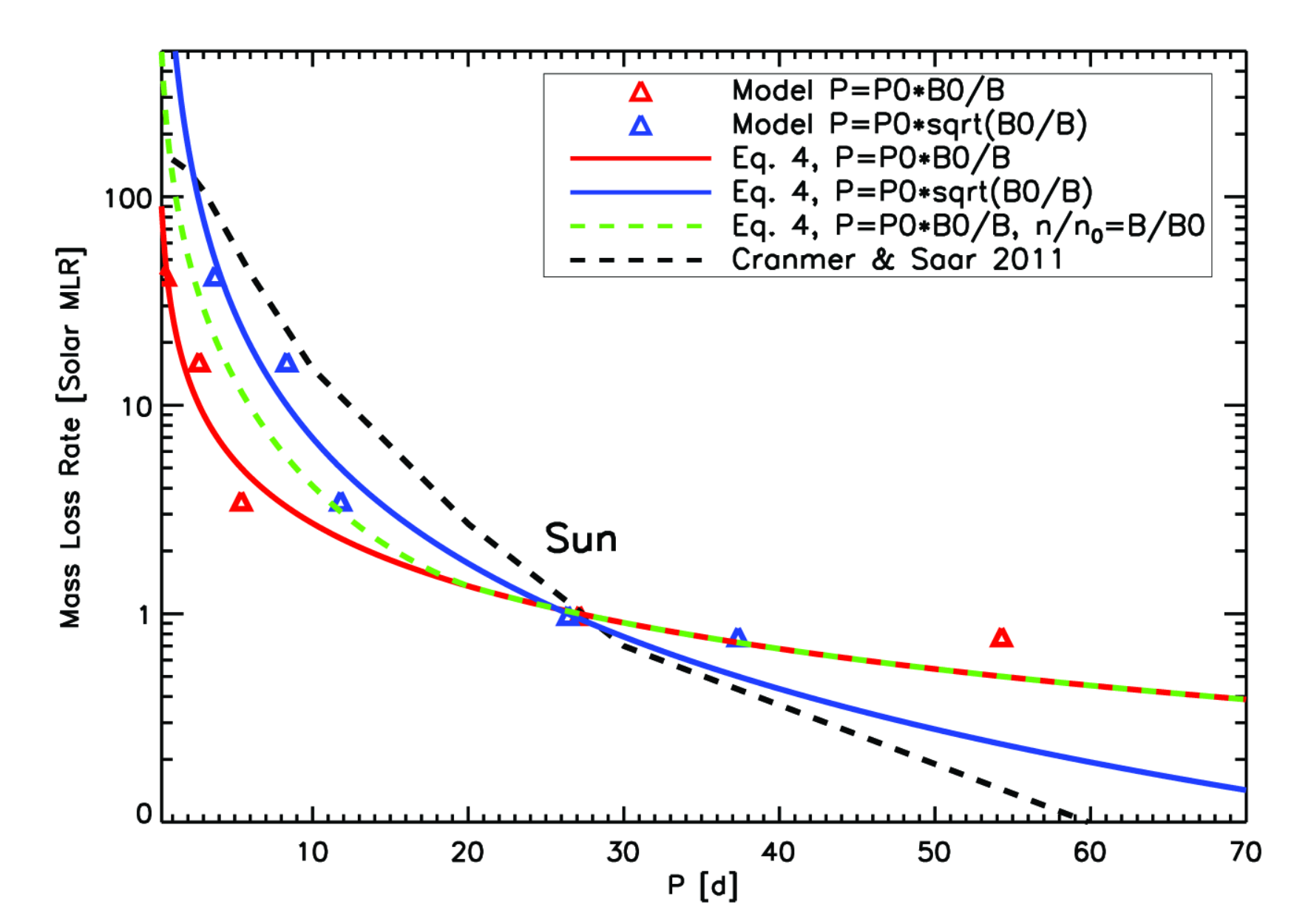}
\caption{Mass loss rate as a function of rotation period for $n=n_\odot$.  Model mass loss rates as a function of rotation period scaled as $B_p\propto \Omega$ (red) and $B_p\propto \Omega^2$ (blue) are marked in triangles, along with the corresponding mass loss rate from Eq.~\ref{MdotScaling}.  The dashed green line represents $B_p\propto \Omega$ and $n\propto \Omega$ scaling, and the dashed black line shows $\dot{M}$ as a function of rotation period from \cite{Cranmer11}.}
\label{fig:f7}
\end{figure*}

Lacking direct measurements, we identify the large-scale stellar dipole fields we have adopted here for our wind model driving with the poloidal field of an $\alpha\Omega$-type dynamo. \citep[e.g.][]{Durney82} used elementary $\alpha-\Omega$ dynamo theory to show that the poloidal magnetic field strength for an unsaturated dynamo should vary approximately linearly with the stellar angular velocity, $B_p\propto \Omega$.  Although the true situation could be considerably more complicated, this relation offers a starting point.  If we assume a linear $\Omega$ dependence of the dipole field strength, adopt a value of 10~G for a solar rotation period of 27~days, and fix the coronal base density at a value of $2\times 10^8$~cm$^{-3}$, we can interpolate among our model solutions and use Eq.~\ref{MdotScaling} to derive the mass loss rate as a function of rotation period.  This curve is illustrated in Figure~\ref{fig:f7}, overplotted with our grid model results for the selected solar parameters.  Also illustrated is a curve assuming a rotation dependence of $B_p\propto\Omega^2$ and the corresponding points from our model, together with the mass loss rate predicted for a solar mass star in the model of \citet{Cranmer11}.  The one remaining free parameter is the base density.   \citet{ivanovataam03} and \citet{HolzwarthJardine07}  assessed evidence from X-ray luminosity vs. rotation to relate the base plasma density to rotation rate adopting $n_\star \propto \Omega_\star^{0.6}$.  While, again, the X-ray emission underlying this assessment is dominated by closed field measurements, some increase in base density with activity level is probably not unreasonable.  For the sake of comparison, we also illustrate in Figure~\ref{fig:f7} a curve corresponding to a case of base density increasing linearly with rotation rate, $n_\star\propto \Omega_\star$.  

We obtain mass loss rates quite similar to those predicted by \citet{Cranmer11} for a fixed base density and a field strength proportional to the square of the rotation velocity.  Perhaps the most interesting region of Figure~\ref{fig:f7} from this perspective is for longer rotation periods than the Sun.  Our model including a rotation-dependent density term, together with a linear dependence of magnetic field strength on rotation velocity, levels off toward a much higher asymptotic mass loss rate than the \citet{Cranmer11} model which continues to decline.  This behavior arises in the density dependent terms in 
Eqn.~\ref{MdotScaling}.   There are unfortunately no observations of mass loss rates for inactive stars with significantly longer rotation periods than the Sun with which to determine whether mass loss continues to decline strongly from the solar rate toward longer rotation periods, or declines much more slowly or levels off. 

While our scaling law for the mass-loss rate covers a range which is similar to that in \citet[e.g.,][]{Wood05}, our mass-loss rate drops much more quickly with rotation rate. The mass-loss rate is still close to 100 times higher than solar for 1 Gyr (10-12 days period) in \cite{Wood05} while it is less than 10 times the solar value in our results. This is quite significant in the context of the mass-loss rate of the young Sun during the early evolution of the Earth and the faint young Sun paradox \citep[see e.g.,][]{Sagan72,Kasting93}. This difference between our results and the mass-loss rate derived from scaling the stellar activity level strengthens the argument made by \cite{Cohen11} that only a weak dependence of the stellar mass-loss on the observed stellar activity level is expected based on solar observations. 

Once a relation between the magnetic field strength and the rotation period is known, 
the angular momentum loss relation in Eqn.~\ref{JdotScaling} can also be used to examine the expected spin-down relation of main-sequence solar-like stars.  Writing the stellar angular momentum as $J=kMR_\star^2\Omega_\star$, where $k$ is a constant depending on the density profile of the stellar interior, and assuming constant mass and radius, we can write $\dot{J}\propto\dot{\Omega_\star}$.  Since $\dot{J}\propto B_p\Omega_\star$, and assuming $B_p\propto\Omega_\star^a$, we have 
\begin{equation}
\frac{d\Omega_\star}{dt}=-\Omega^{(a+1)},
\end{equation}
implying $\Omega_\star \propto t^{-1/a}$.  We therefore retrieve the \citet{Skumanich72} spin-down relation for solar-type stars, $\Omega_\star \propto t^{-1/2}$, for $B_p\propto\Omega^2_\star$.

\section{CONCLUSIONS}
\label{sec:Conclusions}

We have presented a grid of MHD calculations for stellar mass-loss rates and angular momentum loss rates as a function of the stellar dipole field strength, stellar rotation period, and the coronal base density. We find that the loss rates have a simple power-law dependence on the magnetic field strength, in which the slope decreases with an increase of the base density.  The magnetic field strength regulates the wind density and consequently the mass loss rate in the sense that the wind density corresponds to the pressure required to overcome the confining magnetic field.  The resulting wind density has an approximate $B^2$ dependence.  The rotation period itself does not affect the loss rates for rotation periods of more than 2 days, this period corresponding to the building up of a significant azimuthal component of the magnetic field close to the star.  

We derive simple scaling laws for the loss rates as a function of the parameters investigated here. 
These scaling laws generally fit the model results to within a factor of 2.  Based on the fact that a good agreement between the model and observations for the solar case is obtained when using a lower base density, where it does not effect the scaling, and assuming some relation between the rotation period and the dipole magnetic field, the scaling laws for the stellar loss rates are dominated by this rotation - magnetic field relation. 

Despite the success of our simple scaling law in matching the results of the model, we find a deviation between the "solar dipole" case and a real case based on solar minimum observations (where the solar field is close to a dipole). The dipole case produces an Alfv\'en surface that is too small,  and it overestimates the actual solar mass-loss rate by a factor of 3. This comparison with the real Sun implies that the model for stellar fields might require a further investigation with higher complexity, which might include the use of a filling factor for active regions, as well as the distribution of the strength of the small-scale fields. 

For main-sequence solar-like stars, our model results are consistent with the \citet{Skumanich72} relation for spin-down, $\Omega_\star\propto t^{-1/2}$, if the large scale poloidal magnetic field scales with rotation rate as $B_p\propto \Omega^2_\star$.


\acknowledgments
We thank an anonymous referee for her/his useful comments and suggestions. We also thank Vinay Kahsyap and Cecilia Garraffo for useful discussions, and Steve Cranmer for providing the results from his model. This work was supported by {\it Chandra CXC} grant TM2-13001X and SI Grand Challenges grant number 40510298OO5000. We thank the Director of the Unlocking the Mysteries of the Universe Consortium, Christine Jones, for support and encouragement.  JJD was supported by NASA contract NAS8-03060 to the {\it Chandra X-ray Center} and thanks the Director, H.~Tananbaum, for continuing advice and support. Simulation results were obtained using the Space Weather Modeling Framework, developed by the Center for Space Environment Modeling, at the University of Michigan with funding support from NASA ESS, NASA ESTO-CT, NSF KDI, and DoD MURI.




\begin{table}[h!]
\caption{Test Cases shown in Figure~\ref{fig:f3}}
\begin{tabular}{cccc}
\hline
Case & $n_\star\;cm^{-3}$ & $B_\star\;[G]$ & $P_\star\;[d]$ \\
\hline
A &  2e8 & 5 & 0.5 \\ 
B &  2e8 & 5 & 25 \\ 
C &  2e8 & 500 & 0.5 \\ 
D &  2e8 & 500 & 25 \\ 
E &  5e10 & 5 & 0.5 \\ 
F &  5e10 & 5 & 25 \\ 
G &  5e10 & 500 & 0.5 \\ 
H &  5e10 & 500 & 25 \\
\hline
\end{tabular}

\label{table:t1}
\end{table}


\end{document}